\documentclass[a4paper,11pt]{article}

\usepackage{jheppub} 
\usepackage{graphicx}
\usepackage{amsmath,amssymb,amsthm,bm,bigdelim,multirow}
\usepackage{color}
\usepackage{booktabs}
\usepackage{hyperref}
\usepackage{dsfont}
\usepackage{here}
\usepackage{subcaption}
\usepackage{siunitx}

\newcommand{\kslash}{k\kern-1ex /}
\newcommand{\pslash}{p\kern-1ex /}
\newcommand{\qslash}{q\kern-1ex /}
\newcommand{\lslash}{l\kern-1ex /}
\newcommand{\sslash}{s\kern-1ex /}
\newcommand{\Dslash}{D\kern-1.2ex /}

\newcommand{\tr}{{\rm tr}} 

\newcommand{\beqa}{\begin{eqnarray}}
\newcommand{\eeqa}{\end{eqnarray}}
\newcommand{\Tr}{{\rm Tr}}
\newcommand{\be}{\[}
\newcommand{\ee}{\]}
\newcommand{\bd}{\begin{description}}
\newcommand{\ed}{\end{description}}
\newcommand{\la}{\langle}
\newcommand{\ra}{\rangle}
\newcommand{\ben}{\begin{eqnarray}}
\newcommand{\een}{\end{eqnarray}}

\def\lsim{\raise0.3ex\hbox{$<$\kern-0.75em\raise-1.1ex\hbox{$\sim$}}}
\def\gsim{\raise0.3ex\hbox{$>$\kern-0.75em\raise-1.1ex\hbox{$\sim$}}}
\def\simgt{\rlap{\lower 3.5 pt\hbox{$\mathchar \sim$}}\raise 2.0pt \hbox {$>$}}
\def\simlt{\rlap{\lower 3.5 pt\hbox{$\mathchar \sim$}}\raise 2.0pt \hbox {$<$}}

\begin{document}

\title{Critical endpoints of three-dimensional finite density SU(3) spin model with tensor renormalization group}

\author[a]{Xiao Luo},
	\affiliation[a]{Graduate School of Pure and Applied Sciences, University of Tsukuba, Tsukuba, Ibaraki
    305-8571, Japan}
    	\emailAdd{luo@het.ph.tsukuba.ac.jp}

  	\author[b]{Yoshinobu Kuramashi}
  	\affiliation[b]{Center for Computational Sciences, University of Tsukuba, Tsukuba, Ibaraki
    305-8577, Japan}
  	\emailAdd{kuramasi@het.ph.tsukuba.ac.jp}

        \abstract{
We investigate the phase diagram of the three-dimensional SU(3) spin model with finite chemical potential, which is an effective Polyakov loop model for finite density QCD, using the tensor renormalization group method. We successfully determine the location of the critical endpoints being free from the complex action problem in the standard Monte Carlo approach. The critical values of the parameters show the consistency with previous ones obtained by other analytic and numerical methods. 
}
\date{\today}

\preprint{UTHEP-796, UTCCS-P-161}

\maketitle

\section{Introduction}
\label{sec:intro}

The three-dimensional (3$d$) SU(3) spin model with finite chemical potential $\mu\ne 0$ is an effective Polyakov loop model for finite density QCD, which is supposed to describe static quarks in the strong coupling regime. This model has been a popular testbed before exploring finite density QCD since both shares the characteristic features of the spontaneous breaking of the Z(3) center symmetry and the complex action problem at $\mu\ne 0$. So far the Complex Langevin (CL) algorithm~\cite{Karsch:1985cb,Bilic:1987fn,Aarts:2011zn} and the Monte Carlo simulation with a flux representation, flux simulation (FS) in short~\cite{DelgadoMercado:2012gta}, have investigated the phase diagram avoiding the complex action problem. A linked cluster expansion technique has shown that the equation of state and the phase structure are analytically determined~\cite{Kim:2020atu}.

The TRG method~\footnote{In this paper, the ``TRG method" or the ``TRG approach" refers to not only the original numerical algorithm proposed by Levin and Nave \cite{Levin:2006jai} but also its extensions~\cite{PhysRevB.86.045139,Shimizu:2014uva,PhysRevLett.115.180405,Sakai:2017jwp,PhysRevLett.118.110504,Hauru:2017tne,Adachi:2019paf,Kadoh:2019kqk,Akiyama:2020soe,PhysRevB.105.L060402,Akiyama:2022pse}.} is one of the promising tools to investigate the 3$d$ SU(3) spin model at finite density. It has been successfully applied to various types of models with the sign problem or the complex action problem~\cite{Shimizu:2014uva,Shimizu:2014fsa,Kawauchi:2016xng,Kawauchi:2017dnj,Yang:2015rra,Shimizu:2017onf,Takeda:2014vwa,Kadoh:2018hqq,Kadoh:2019ube,Kuramashi:2019cgs,Akiyama:2020ntf,Akiyama:2020soe,Nakayama:2021iyp,Akiyama:2023hvt,Akiyama:2024qer}. In this paper we investigate the phase structure of the model and determine the critical endpoints. Our results for the critical couplings are compared with the previous ones obtained with the flux simulation~\cite{DelgadoMercado:2012gta} and the analytical cluster expansion method~\cite{Kim:2020atu}.

This paper is organized as follows. In Sec.~\ref{sec:method}, we define the action of the 3$d$ SU(3) spin model with finite chemical potential and give the tensor network representation. 
In Sec.~\ref{sec:results} we determine the critical endpoints as a function of the coupling parameters and  the results are compared with the previous ones obtained by different approaches.
Section~\ref{sec:summary} is devoted to summary and outlook.

\section{Formulation and numerical algorithm}
\label{sec:method}

We consider the SU(3) spin model at finite density on a 3$d$ lattice $\Lambda_3=\{(n_1,n_2,n_3)\ \vert n_{1,2,3}=1,\dots,L\}$ whose volume is $V=L^3$. The lattice spacing $a$ is set to $a=1$ unless necessary. The action is defined as~\cite{Karsch:1985cb}
\begin{equation}
  S[U] = -\sum_{n\in\Lambda_3}\left\{\sum_{\nu=1}^{3}\beta \left[P(n)P^*(n+{\hat{\nu}})+P^*(n)P(n+{\hat{\nu}})\right]+\kappa\left[e^\mu P(n)+e^{-\mu}P^*(n)\right]\right\}
  \label{eq:action}
\end{equation}
with $\hat{\nu}$ the unit vectors in the $\nu$ direction and $\mu$ the chemical potential. The fields $P(n)$ represent the traced SU(3) matrices $P(n)=\tr U(n)$ with $U(n)\in \rm{SU(3)}$. This model describes an effective theory of 4$d$ finite density QCD at the strong coupling with heavy quarks, where $P(n)$ represent the Polyakov loops and $(\beta, \kappa)$ are effective couplings related to the gauge coupling and the quark mass, respectively. 
The grand canonical partition function of the model is expressed as
\be
Z(\beta,\kappa,\mu)=\int \prod_{n\in\Lambda_3} dU(n) e^{-S[U]}
\ee
with $dU(n)$ the Haar measure at the site $n$.
Here we note that the $U(n)$ matrices can be diagonalized simultaneously over all sites of the lattice and the eigenvalues are given by $\exp(i\theta)$, $\exp(i\phi)$ and $\exp(-i(\theta+\phi))$. The action and the partition function is rewritten as follows:
\ben
Z(\beta,\kappa,\mu) &=& \int {\cal D}\Omega e^{-S[P]} \label{eq:partitionfunction}, \\
{\cal D}\Omega &=& \prod_{n\in\Lambda_3}\sin^2\left(\frac{\theta_n-\phi_n}{2}\right)\sin^2\left(\frac{2\theta_n+\phi_n}{2}\right)\sin^2\left(\frac{\theta_n+2\phi_n}{2}\right)d\theta_n d\phi_n
\een
with
\be
\begin{array}{c}
\Omega=(\theta,\phi) \quad, ~\theta \in (0,2\pi],~\phi \in (0,2\pi], 
\end{array}
\ee
where $P(n)$ in the action $S[P]$ is given by
\be
P(n)=\exp(i\theta_n)+\exp(i\phi_n)+\exp(-i(\theta_n+\phi_n)).
\ee
In case of $\kappa=0$ the action has the global Z(3) symmetry
with Z(3) the center of SU(3).

We discretize the integration (\ref{eq:partitionfunction}) with the Gauss-Legendre quadrature~\cite{Kuramashi:2019cgs,Akiyama:2020ntf} after changing the integration variables:
\ben
-1 \le \alpha&=&\frac{1}{\pi}\left(\theta-\pi \right)\le 1, \\
-1 \le \beta&=&\frac{1}{\pi}\left(\phi-\pi \right)\le 1. 
\een
Then we obtain the discretized partition function
\begin{align}
	Z =& \sum_{ \{\Omega_1\},\cdots,\{\Omega_V\}} \left[ \prod_{n=1}^{V}  \sin^2\left(\frac{\theta(\alpha_{a_n})-\phi(\beta_{b_n})}{2}\right)\sin^2\left(\frac{2\theta(\alpha_{a_n})+\phi(\beta_{b_n})}{2}\right) \right. \nonumber \\
	&  \qquad\qquad\qquad \times \left. \sin^2\left(\frac{\theta(\alpha_{a_n})+2\phi(\beta_{b_n})}{2}\right) w_{a_n} w_{b_n} \right] \prod_{\nu=1}^3 M_{\Omega_n,\Omega_{n+\hat{\nu}}},
\end{align}
with $\Omega_n=(\theta(\alpha_{a_n}),\phi(\beta_{b_n}))\equiv (a_n,b_n)$, where $\alpha_{a_n}$ and $\beta_{b_n}$ are $a$- and $b$-th roots of the $K$-th Legendre polynomial $P_{K}(s)$ on the site $n$, respectively. $\sum_{ \{\Omega_n\}}$ denotes $\sum_{a_n=1}^{K}\sum_{b_n=1}^{K}$.
$M$ is a 4-legs tensor defined by
\begin{align}
	M_{a_n,b_n,a_{n+\hat{\nu}}, b_{n+\hat{\nu}}}  
	=& \exp\left\{ \beta\left[P(a_n, b_n)P^*(a_{n+{\hat{\nu}}}, b_{n+{\hat{\nu}}})+P^*(a_n, b_n)P(a_{n+{\hat{\nu}}}, b_{n+{\hat{\nu}}})\right] \right. \nonumber \\
	& \qquad + \left.\kappa\left[e^\mu P(a_n,b_n)+e^{-\mu}P^*(a_n,b_n)\right] \right\}~.
\end{align} 
The weight factor $w$ of the Gauss-Legendre quadrature is defined as
\begin{equation}
	w_{a_n} = \frac{2(1-{\alpha^2_{a_n}})}{K^2P^2_{K-1}({\alpha_{a_n}})},\quad
	w_{b_n} = \frac{2(1-{\beta^2_{b_n}})}{K^2P^2_{K-1}({\beta_{b_n}})}.
\end{equation}
Throughout this paper we employ $K=100$. 
After performing the singular value decomposition (SVD) on $M$:
\begin{equation}
	M_{a_n,b_n,a_{n+\hat{\nu}}, b_{n+\hat{\nu}}} \simeq \sum_{i_n=1}^{D} U_{a_n,b_n, i_n} \sigma_{i_n} V^\dagger_{i_n,a_{n+\hat{\nu}}, b_{n+\hat{\nu}}},
\end{equation}
where $U$ and $V$ denotes unitary matrices and $\sigma$ is a diagonal matrix with the singular values of $M$ in the descending order.
We can obtain the tensor network representation of the SU(3) spin model on the site $n\in\Lambda_{3}$
\begin{align}
	T_{x_n, x'_n, y_n, y'_n, z_n, z'_n} &= \sqrt{\sigma_{x_n} \sigma_{x'_n} \sigma_{y_n} \sigma_{y'_n} \sigma_{z_n} \sigma_{z'_n}} \nonumber \\
	&\quad \times \sum_{a_n, b_n} \left\{ \sin^2\left(\frac{\theta(\alpha_{a_n})-\phi(\beta_{b_n})}{2}\right)\sin^2\left(\frac{2\theta(\alpha_{a_n})+\phi(\beta_{b_n})}{2}\right) \right. \nonumber \\
	&\quad \times \sin^2\left(\frac{\theta(\alpha_{a_n})+2\phi(\beta_{b_n})}{2}\right) w_{a_n} w_{b_n}\nonumber \\
	&\quad \left. \times V^\dagger_{x_n,a_n,b_n} U_{a_n,b_n, x'_n} V^\dagger_{y_n,a_n,b_n} U_{a_n,b_n, y'_n} V^\dagger_{z_n,a_n,b_n} U_{a_n,b_n, z'_n} \right\} , 
\end{align} 
where the bond dimension of tensor $T$ is given by  $D$, which controls the numerical precision in the TRG method. The tensor network representation of partition function is given by
\begin{equation}
  Z \simeq \sum_{x_0 x'_0 y_0 y'_0 z_0 z'_0 \cdots} \prod_{n \in \Lambda_{3}} T_{x_n x'_n y_n y'_n z_n z'_n} = \Tr \left[T \cdots T\right]~.
  \label{eq:Z_TN}
\end{equation}
In order to evaluate $Z$ we employ the Anisotropic Tensor Renormalization Group (ATRG) algorithm~\cite{Adachi:2019paf}.

\section{Numerical results}
\label{sec:results}

Before discussing the determination of the critical endpoints it is worthwhile to make a consistency check of the values of $\la P\ra$ and  $\la P^*\ra$ with other approaches. In Table~\ref{tab:cmparison_P} we compare the results of $\la P\ra$ and $\la P^*\ra$ in the finite $\mu$ case at $(\beta,\kappa)=(0.125,0.02)$ with those obtained from the CL and FS methods~\cite{Aarts:2011zn,DelgadoMercado:2012gta}. Note that our results are evaluated on an $8^3$ lattice due to the coarse graining procedure, while the CL and FS results are on a $10^3$ lattice. The results show an good agreement within the error bars.     

\begin{table}[htb]
	\caption{Comparison of $\la P\ra$ and  $\la P^*\ra$ at $(\beta,\kappa)=(0.125,0.02)$ with the choices of $\mu=1.0$ and 3.0. }
	\label{tab:cmparison_P}
	\begin{center}
	  	\begin{tabular}{|cc|ll|ll|}\hline
          	&&\multicolumn{2}{|c|}{$\mu=1.0$}&\multicolumn{2}{|c|}{$\mu=3.0$}  \\ \hline
		 & lattice size & $\la P\ra$ & $\la P^*\ra$ & $\la P\ra$ & $\la P^*\ra$ \\ \hline
Complex Langevin~\cite{Aarts:2011zn} & $10^3$ & 0.2419(19) & 0.3605(13) & 1.70615(27) & 1.74590(24) \\
Flux simulation~\cite{DelgadoMercado:2012gta} & $10^3$ & 0.2416(11) & 0.3604(13) & 1.70627(14) & 1.74683(17) 	\\
This work ($D=96$) & $8^3$ & 0.242620 & 0.358987 & 1.706390 & 1.746148	 \\ \hline
		\end{tabular}
	\end{center}
\end{table}

\subsection{Critical endpoints} 
\label{subsec:results}

The previous studies~\cite{Karsch:1985cb,Bilic:1987fn,Aarts:2011zn,DelgadoMercado:2012gta,Kim:2020atu} suggest that the phase structure of the model is rather simple: There exists a critical endline in the three dimensional parameter space $(\beta,\kappa,\mu)$, which forms a boundary of the first-order phase transition surface. In this work we focus on two cases fixed at $\mu=0$ and $\kappa=0.005$. These specific cases were investigated in detail by using the flux representation technique~\cite{DelgadoMercado:2012gta} and the analytical cluster expansion method~\cite{Kim:2020atu}.

We closely follow the procedure employed in Refs.~\cite{Akiyama:2022eip,Akiyama:2023hvt} to determine the critical endpoints. We measure the expectation value of the field $P$ defined by
\be
\langle P \rangle=\frac{1}{V}\frac{\partial \ln Z}{\partial \eta}
\ee
with $\eta=\kappa e^{\mu}$ to detect the first-order phase transition using the impurity tensor method.

\begin{figure}[htbp]
	\centering
	\includegraphics[width=0.8\hsize]{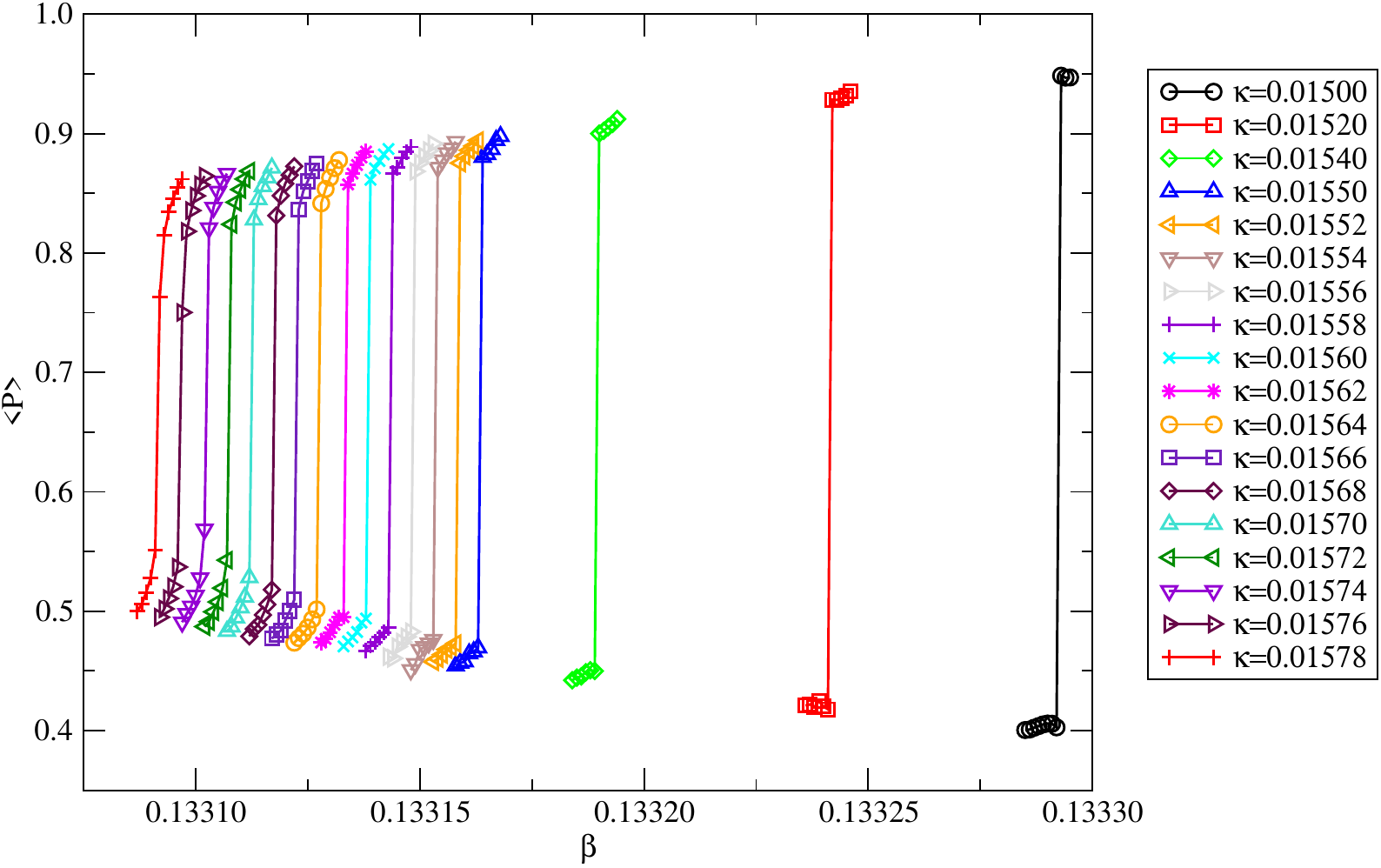}
	\caption{$\beta$ dependence of $\langle P\rangle$ at $\mu=0$ for $\kappa\in[0.01500,0.01578]$ with $D=88$ on a $1024^3$ lattice.}
  	\label{fig:P_mu0}
\end{figure}

\begin{figure}[htbp]
	\centering
	\includegraphics[width=0.8\hsize]{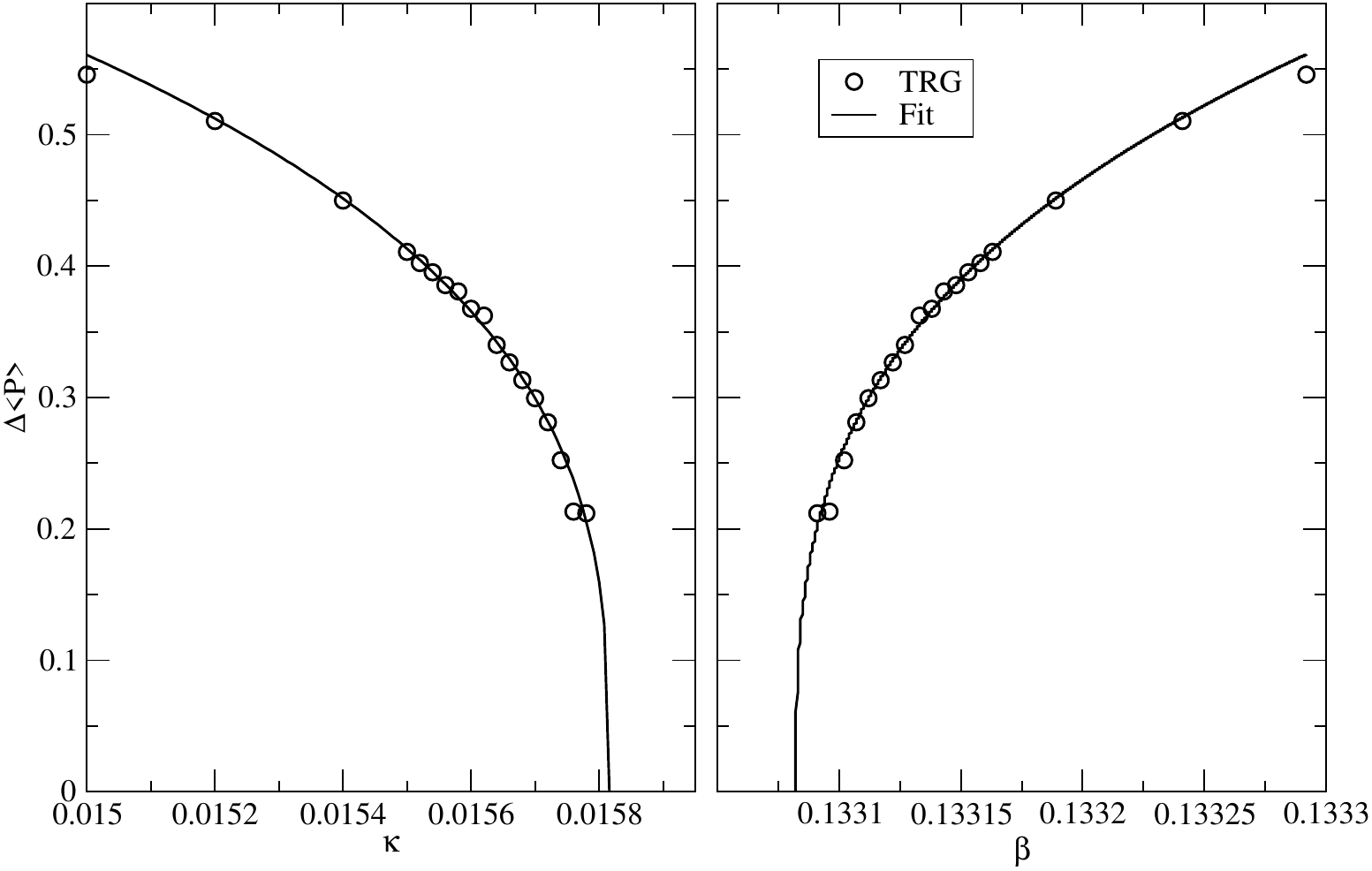}
	\caption{Fit of $\Delta \langle P\rangle$ at $\mu=0$ with $D=88$ as a function of $\kappa$ (left) and $\beta$ (right).}
  	\label{fig:fit_mu0}
\end{figure}

We first present the results in the $\mu=0$ case. Figure~\ref{fig:P_mu0} shows the $\beta$ dependence of $\langle P\rangle$  at $D=88$ as a representative case with the several choices of $\kappa\in [0.01500,0.01578]$, where the clear gap of $\langle P\rangle$ is observed at a certain value of $\kappa$ for $\kappa\in[0.01500,0.01572]$. The value of the gap of $\langle P\rangle$, which is denoted by $\Delta \langle P\rangle$, is evaluated with $\langle P\rangle(\beta=\beta_+)-\langle P\rangle(\beta=\beta_-)$, where $\beta_+$ and $\beta_-$ are chosen from different phases. We set $\vert \beta_+-\beta_-\vert=10^{-6}$ for $\kappa\in [0.01500,0.01578]$. In order to determine the critical endpoint $(\beta_{\rm c}(D),\kappa_{\rm c}(D))$ at $\mu=0$ we make a simultaneous fit of $\Delta \langle P\rangle$ in the region of $\kappa\in [0.01540,0.01572]$ assuming the functions of $\Delta \langle P\rangle=A(\beta-\beta_{\rm c}(D))^p$ and $\Delta \langle P\rangle=B(\kappa_{\rm c}(D)-\kappa)^p$ with $A$, $B$, $\beta_{\rm c}(D)$, $\kappa_{\rm c}(D)$ and $p$ the free parameters. The fit results are depicted in Fig.~\ref{fig:fit_mu0} and their numerical values are presented in Table~\ref{tab:fit_mu0}.

\begin{table}[htb]
	\caption{Fit results for $\Delta \langle P\rangle$ at $\mu=0$. $[\kappa_{\rm min},\kappa_{\rm max}]$ denotes the fit range. }
	\label{tab:fit_mu0}
	\begin{center}
	  	\begin{tabular}{|ccSllll|}\hline
          	\multicolumn{7}{|c|}{$\mu=0$}  \\ \hline
		$D$ & $[\kappa_{\rm min},\kappa_{\rm max}]$ & $A$ & $\beta_{\rm c}{(D)}$ & $p$ & $B$ & $\kappa_{\rm c}{(D)}$ \\ \hline
		36 & $[0.01550,0.01598]$ & 8.1(7) & 0.133016(2) & 0.325(10) & 5.2(4) & 0.016119(9) \\
		40 & $[0.01550,0.01586]$ & 7.9(13) & 0.133013(4)& 0.315(19) & 5.1(7) & 0.015973(14) \\
		44 & $[0.01550,0.01582]$ & 8.9(14) & 0.132997(5)& 0.333(19) & 5.7(8) & 0.016040(20)\\
		48 & $[0.01550,0.01582]$ & 8.0(8) & 0.133054(2) & 0.322(12) & 5.2(4) & 0.015942(8) \\
		52 & $[0.01550,0.01584]$ & 7.9(10)& 0.133042(3) & 0.316(15) & 5.2(6) & 0.015986(12) \\
        56 & $[0.01550,0.01584]$ & 8.4(9) & 0.133957(3) & 0.324(13) & 5.4(5) & 0.015976(10) \\
        64 & $[0.01540,0.01572]$ & 8.7(9) & 0.133077(2) & 0.325(12) & 5.6(5) & 0.015810(7) \\
		68 & $[0.01540,0.01570]$ & 10.6(17) & 0.133087(2) & 0.347(18) & 6.6(9) & 0.015781(9) \\
        72 & $[0.01540,0.01570]$ & 7.4(21) & 0.133095(4) & 0.306(31) & 4.9(12) & 0.015755(14) \\
		76 & $[0.01540,0.01568]$ & 8.4(15) & 0.133082(3) & 0.320(20) & 5.4(8) & 0.015795(13) \\
        80 & $[0.01540,0.01568]$ & 10.3(17) & 0.133085(3) & 0.341(19) & 6.5(9) & 0.015788(11) \\
		84 & $[0.01540,0.01572]$ & 8.5(5) & 0.133083(1) & 0.323(7) & 5.5(3) & 0.015811(4) \\
        88 & $[0.01540,0.01572]$ & 8.6(11) & 0.133082(2) & 0.322(14) & 5.5(6) & 0.015816(8) \\
		92 & $[0.01550,0.01572]$ & 8.5(7) & 0.133100(1) & 0.320(8) & 5.5(4) & 0.015753(2) \\
        96 & $[0.01552,0.01572]$ & 8.9(6) & 0.1331093(4) & 0.324(7) & 5.7(3) & 0.015742(2) \\ \hline
		\end{tabular}
	\end{center}
\end{table}

\begin{figure}[htbp]
	\centering
	\includegraphics[width=1.0\hsize]{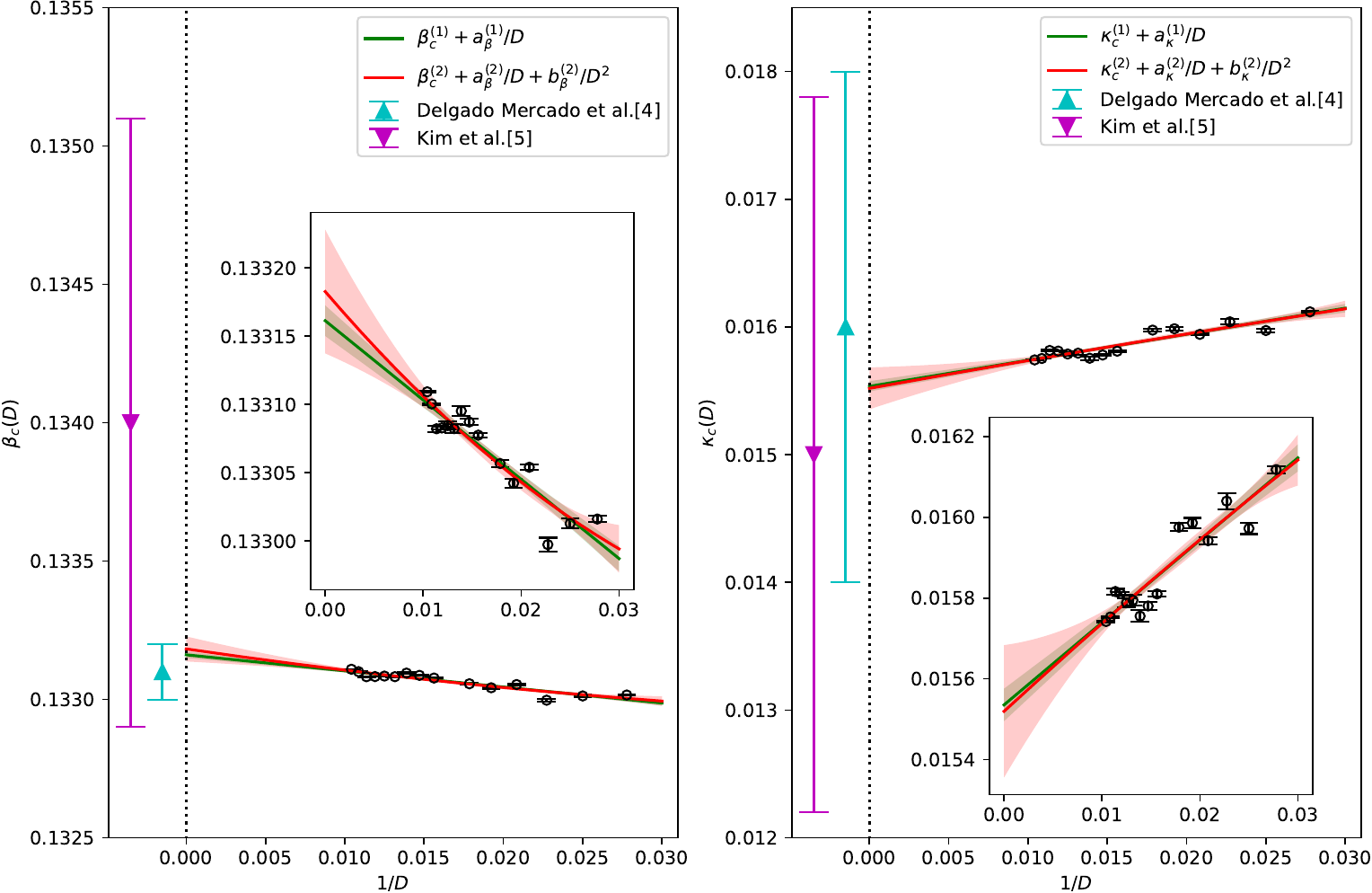}
  \caption{(Left) $\beta_{\rm c}(D)$ as a function of $1/D$. Green and red curves represent the fitting results with the functions $\beta_{\rm c}(D)=\beta_{\rm c}^{(1)}(\infty)+a_\beta^{(1)} /D$ and $\beta_{\rm c}(D)=\beta_{\rm c}^{(2)}(\infty)+a_\beta^{(2)} /D+b_\beta^{(2)} /D^2$. Shaded areas denote the error bands. (Right) Same as the left panel for $\kappa_{\rm c}(D)$.}
  	\label{fig:extrap_mu0}
\end{figure}

In order to extrapolate the results of $\beta_{\rm c}(D)$ and $\kappa_{\rm c}(D)$ to the limit $D\rightarrow \infty$, we repeat the same calculation changing the value of $D$. The fit results for $\Delta \langle P\rangle$ are summarized in Table~\ref{tab:fit_mu0}. In the left panel of Fig.~\ref{fig:extrap_mu0} we plot $\beta_{\rm c}(D)$ as a function of $1/D$. The data show upward trend, while staggering, as $1/D$ decreases. The red curve shows the fitting result with the quadratic function $\beta_{\rm c}(D)=\beta_{\rm c}^{(2)}(\infty)+a_\beta^{(2)} /D+b_\beta^{(2)} /D^2$ yielding $\beta_{\rm c}^{(2)}(\infty)=0.133183(44)$, where the shaded area denotes the error band. The extrapolated result is consistent with the previous ones obtained with the flux simulation~\cite{DelgadoMercado:2012gta} and the analytical cluster expansion method~\cite{Kim:2020atu}, whose numerical values are listed in Table~\ref{tab:comparison}. For comparison we also show the fitting result with the linear function denoted by the green line. The inset magnifies the vertical scale to clarify the difference between the linear and quadratic results. We observe that the former is within the error band in the latter.
The same procedure is employed to obtain the value of $\kappa_{\rm c}^{(2)}(\infty)=0.01552(16)$. The fit results for $\kappa_{\rm c}(D)$ are presented in the right panel of Fig.~\ref{fig:extrap_mu0}. In this case the linear and quadratic results are almost degenerate. In Table~\ref{tab:fit_mu0} we observe that the values of $p$ scatter around $p=0.326$ expected for the 3$d$ Ising universality class.

\begin{figure}[htbp]
	\centering
	\includegraphics[width=0.8\hsize]{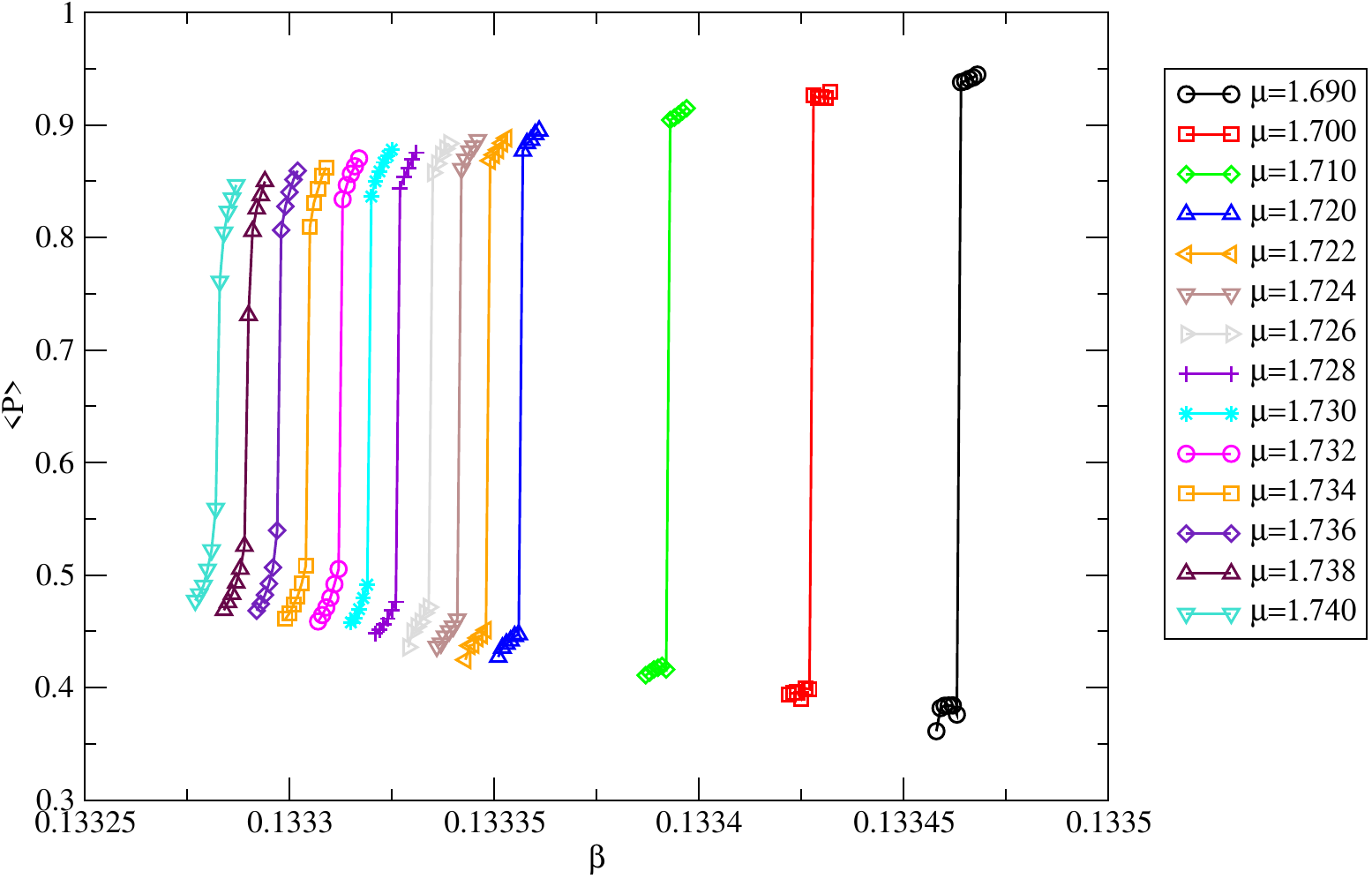}
	\caption{$\beta$ dependence of $\langle P\rangle$ at $\kappa=0.005$ for $\mu\in[1.690,1.740]$ with $D=88$ on a $1024^3$ lattice.}
  	\label{fig:P_k0005}
\end{figure}

\begin{figure}[htbp]
	\centering
	\includegraphics[width=0.8\hsize]{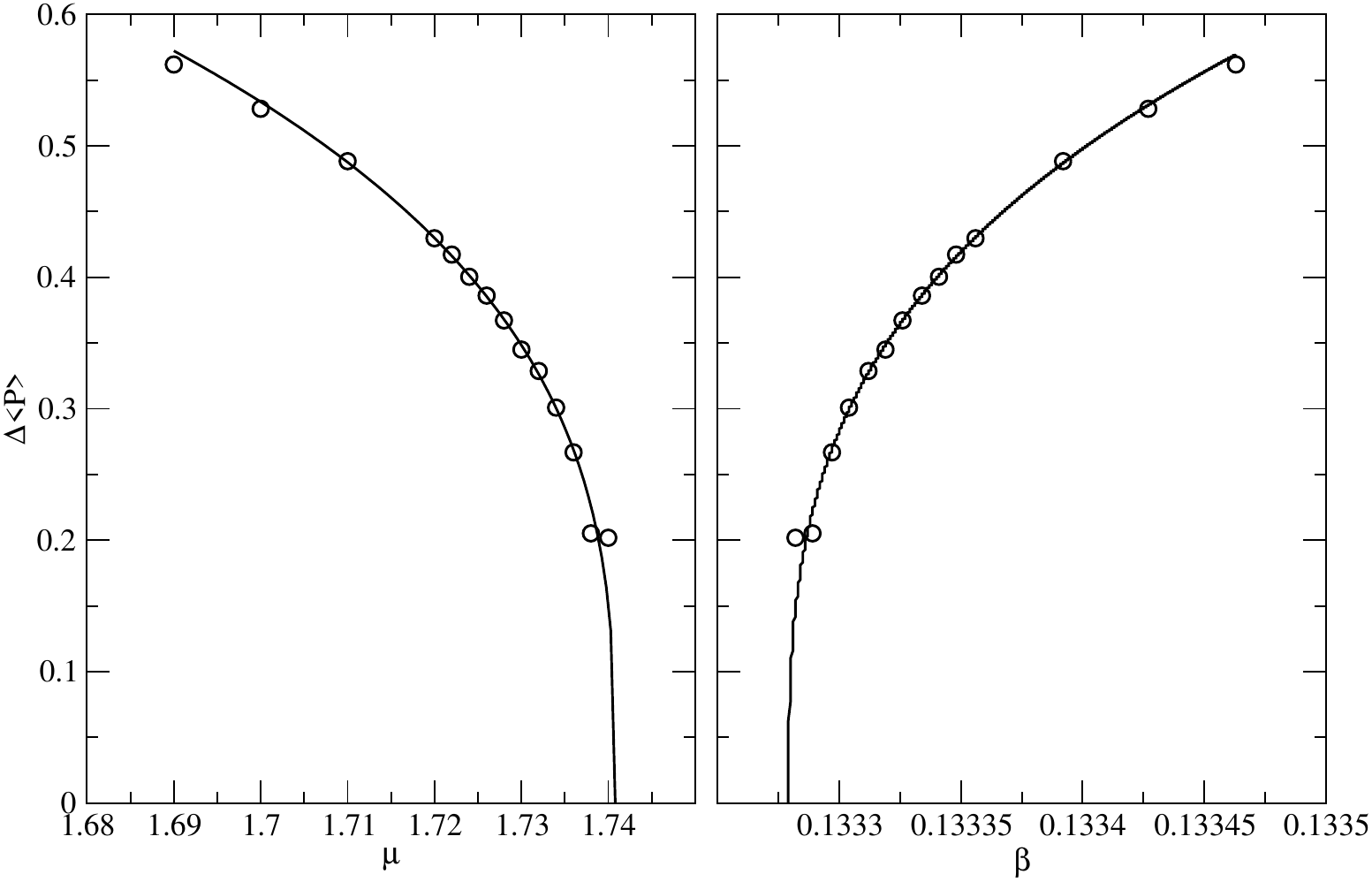}
	\caption{Fit of $\Delta \langle P\rangle$ at $\kappa=0.005$ with $D=88$ as a function of $\mu$ (left) and $\beta$ (right).}
  	\label{fig:fit_k0005}
\end{figure}

Now we move on to the $\kappa=0.005$ case. The $\beta$ dependence of $\langle P\rangle$ for $\mu\in [1.690,1.740]$ is shown in Fig.~\ref{fig:P_k0005}, where we choose $D=88$ as a representative case. We observe clear gaps of $\langle P\rangle$ with $\mu\in [1.690,1.734]$, though it is difficult to identify such a gap at $\mu=1.738$. As in the $\mu=0$ case, we determine $(\beta_{\rm c}(D),\mu_{\rm c}(D))$ by a simultaneous fit of $\Delta \langle P\rangle$ for $\mu\in [1.710,1.734]$ employing the functions of $\Delta \langle P\rangle=A(\beta-\beta_{\rm c}(D))^p$ and $\Delta \langle P\rangle=B(\mu_{\rm c}(D)-\mu)^p$ with $A$, $B$, $\beta_{\rm c}(D)$, $\mu_{\rm c}(D)$ and $p$ the free parameters. The fit results are  presented in Fig.~\ref{fig:fit_k0005} together with the numerical values in Table~\ref{tab:fit_k0005}. Note that we measure $\Delta \langle P\rangle=\langle P\rangle(\beta=\beta_+)-\langle P\rangle(\beta=\beta_-)$ with $\vert \beta_+-\beta_-\vert =10^{-6}$. As in the case of $\mu=0$  the results for the critical exponent $p$ in Table~\ref{tab:fit_k0005} show rough consistency with $p=0.326$ in the 3$d$ Ising universality class.

\begin{table}[htb]
	\caption{Comparison of critical endpoints in the $\mu=0$ and $\kappa=0.005$ cases.}
	\label{tab:comparison}
	\begin{center}
	\begin{tabular}{|c|ll|ll|}\hline
          &\multicolumn{2}{|c|}{$\mu=0$}&\multicolumn{2}{|c|}{$\kappa=0.005$}  \\ \hline
           & \multicolumn{1}{c}{$\beta_{\rm c}$} & \multicolumn{1}{c|}{$\kappa_{\rm c}$} & \multicolumn{1}{c}{$\beta_{\rm c}$} & \multicolumn{1}{c|}{$\mu_{\rm c}$} \\ \hline
	Delgado Mercado {\it et al.}~\cite{DelgadoMercado:2012gta} & 0.1331(1) & 0.016(2) & 0.134(1) & 1.53(10) \\ 
	Kim {\it et al.}~\cite{Kim:2020atu} & 0.1340(11) & 0.0150(28) & 0.1346(19) & 1.61(27) \\ 
	This work & 0.133183(44) & 0.01552(16) & 0.133393(45) & 1.716(12) \\ \hline
		\end{tabular}
	\end{center}
\end{table}

\begin{table}[htb]
	\caption{Fit results for $\Delta \langle P\rangle$ at $\kappa=0.005$. $[\mu_{\rm min},\mu_{\rm max}]$ denotes the fit range.}
	\label{tab:fit_k0005}
	\begin{center}
	  	\begin{tabular}{|cclllll|}\hline
          	\multicolumn{7}{|c|}{$\kappa=0.005$}  \\ \hline
		$D$ & $[\mu_{\rm min},\mu_{\rm max}]$ & $A$ & $\beta_{\rm c}(D)$ & $p$ & $B$ & $\mu_{\rm c}(D)$ \\ \hline
	36 & $[1.720,1.750]$ & 8.8(12) & 0.133226(3) & 0.324(16) & 1.43(7) & 1.7591(9) \\
	40 & $[1.722,1.750]$ & 9.5(9) & 0.133205(1) & 0.332(10) & 1.49(5) & 1.7532(3) \\
	44 & $[1.720,1.742]$ & 9.6(16) & 0.133194(5) & 0.336(20) & 1.46(8) & 1.7566(14) \\
	48 & $[1.720,1.742]$ & 8.9(6) & 0.133237(2) & 0.329(8) & 1.42(3) & 1.7537(5) \\
	52 & $[1.720,1.742]$ & 8.8(5) & 0.133236(1) & 0.322(6) & 1.46(3) & 1.7538(2) \\
    56 & $[1.720,1.746]$ & 8.7(7) & 0.133244(2) & 0.322(10) & 1.44(4) & 1.7549(5) \\
    64 & $[1.710,1.732]$ & 9.5(8) & 0.133269(2) & 0.330(9) & 1.48(4) & 1.7423(5) \\
    68 & $[1.710,1.732]$ & 9.2(8) & 0.133281(1) & 0.325(10) & 1.48(5) & 1.7392(4) \\
	72 & $[1.710,1.734]$ & 8.9(8) & 0.133283(1) & 0.317(10) & 1.49(5) & 1.7390(4) \\
	76 & $[1.710,1.732]$ & 10.2(11) & 0.133277(2) & 0.335(13) & 1.57(6) & 1.7401(6) \\
    80 & $[1.710,1.738]$ & 9.5(9) & 0.133266(2) & 0.328(10) & 1.51(5) & 1.7438(4) \\
	84 & $[1.710,1.732]$ & 8.0(18) & 0.133278(5) & 0.308(26) & 1.42(11) & 1.7412(14) \\
    88 & $[1.710,1.734]$ & 9.0(6) & 0.133279(1) & 0.320(8) & 1.49(4) & 1.7408(3) \\
    92 & $[1.710,1.730]$ & 9.3(9) & 0.133295(2) & 0.323(12) & 1.51(6) & 1.7373(5) \\
    96 & $[1.710,1.730]$ & 9.0(10) & 0.133286(3) & 0.321(13) & 1.48(6) & 1.7411(8)  \\ \hline
		\end{tabular}
	\end{center}
\end{table}

\begin{figure}[htb]
	\centering
	\includegraphics[width=1.0\hsize]{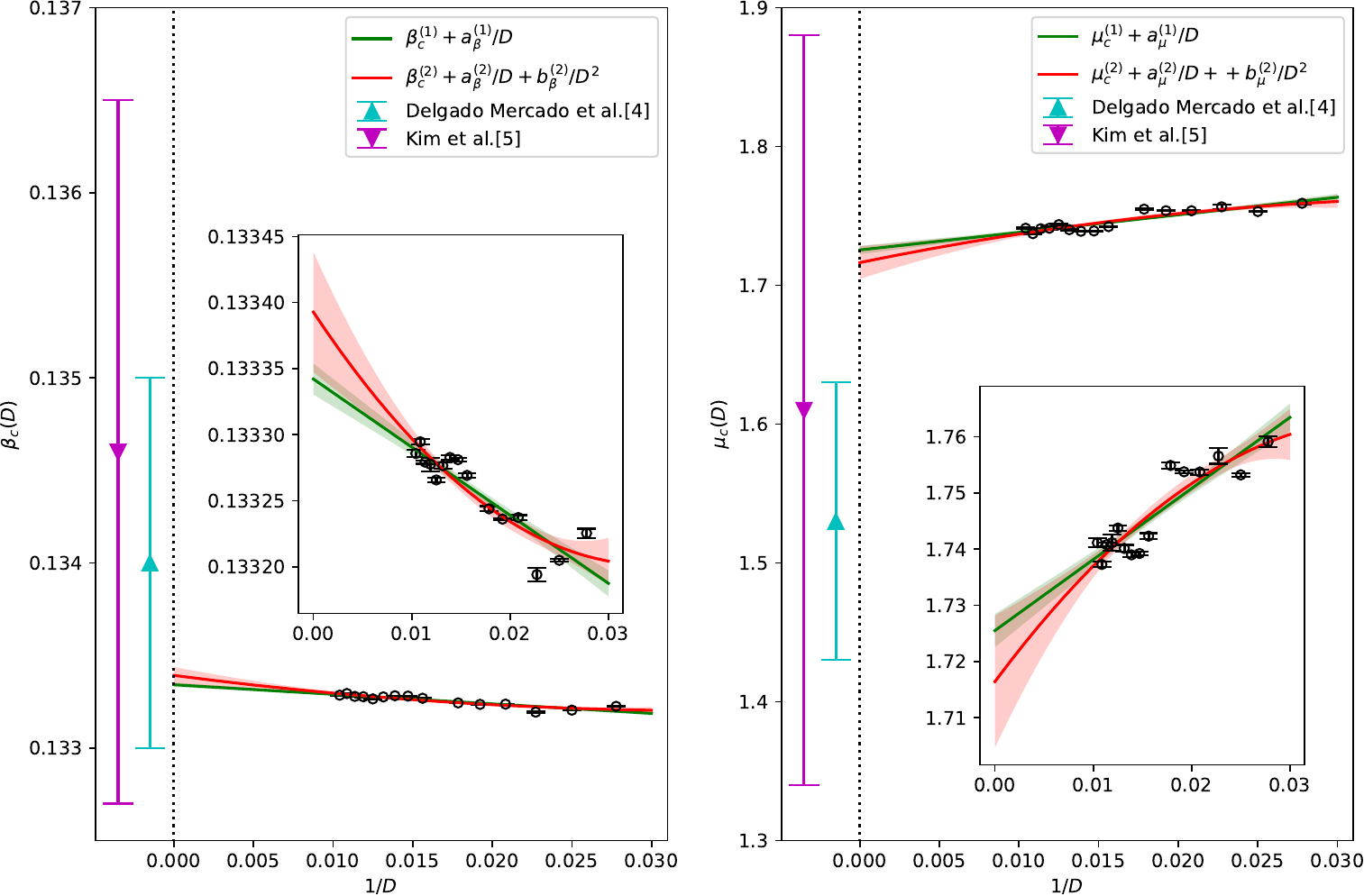}
  \caption{(Left) $\beta_{\rm c}(D)$ as a function of $1/D$. Green and red curves represent the fitting results with the functions $\beta_{\rm c}(D)=\beta_{\rm c}^{(1)}(\infty)+a_\beta^{(1)} /D$ and $\beta_{\rm c}(D)=\beta_{\rm c}^{(2)}(\infty)+a_\beta^{(2)} /D+b_\beta^{(2)} /D^2$. Shaded areas denote the error bands. (Right) Same as the left panel for $\mu_{\rm c}(D)$.}
  	\label{fig:extrap_k0005}
\end{figure}

We perform the fit of $\Delta \langle P\rangle$ with other choices of $D$. The results of the fit parameters including the critical endpoint $(\beta_{\rm c}(D),\mu_{\rm c}(D))$ are listed in Table~\ref{tab:fit_k0005}. As in the $\mu=0$ case we obtain $\beta_{\rm c}^{(2)}(\infty)$ and $\mu_{\rm c}^{(2)}(\infty)$ employing the quadratic extrapolation. The left panel of Fig.~\ref{fig:extrap_k0005} shows $1/D$ dependence of $\beta_{\rm c}(D)$ together with the fit result with the function $\beta_{\rm c}(D)=\beta_{\rm c}^{(2)}(\infty)+a_\beta^{(2)} /D+b_\beta^{(2)} /D^2$.  The right panel is for $\mu_{\rm c}(D)$. We obtain $\beta_{\rm c}^{(2)}(\infty)=0.133393(45)$ and $\mu_{\rm c}^{(2)}(\infty)=1.716(12)$ as listed in Table~\ref{tab:comparison}. While they agree with ones with the analytical cluster expansion method~\cite{DelgadoMercado:2012gta}, the value of $\mu_{\rm c}^{(2)}(\infty)$ is slightly beyond the error bar of the result of the flux simulation~\cite{Kim:2020atu}.

\section{Summary and outlook} 
\label{sec:summary}

We have investigated the phase structure of the 3$d$ SU(3) spin model with the TRG method focusing on the critical endpoints at the $\mu=0$ and $\kappa=0.005$ cases.  We find $(\beta_{\rm c},\kappa_{\rm c})=(0.133183(44) ,0.01552(16))$ for $\mu=0$ and $(\beta_{\rm c},\mu_{\rm c})=(0.133393(45), 1.716(12))$ for $\kappa=0.005$ in the $D\rightarrow \infty$ limit, which are consistent with the previous results obtained by the flux representation technique and the analytical cluster expansion method. The critical exponents also show a consistency with the value expected from the 3$d$ Ising universality class. This is an encouraging result as a stepping stone toward the investigation of the phase diagram of 4$d$ finite density QCD.   

\begin{acknowledgments}
  Numerical calculation for the present work was carried out with the supercomputers Cygnus and Pegasus under the Multidisciplinary Cooperative Research Program of Center for Computational Sciences, University of Tsukuba. We also used the supercomputer Fugaku provided by RIKEN through the HPCI System Research Project (Project ID: hp230247).
This work is supported in part by Grants-in-Aid for Scientific Research from the Ministry of Education, Culture, Sports, Science and Technology (MEXT) (Nos. 24H00214, 24H00940).

\end{acknowledgments}



\bibliographystyle{JHEP}
\bibliography{bib/formulation,bib/algorithm,bib/discrete,bib/grassmann,bib/continuous,bib/gauge,bib/review,bib/for_this_paper}

\end{document}